\documentclass{article}
\usepackage{spconf,amsmath,graphicx}
\usepackage{tipa}
\usepackage{tabularx}
\usepackage{booktabs}
\usepackage{array}
\usepackage{color}
\usepackage{graphics}

\title{How phonemes contribute to deep speaker models?}
\name{Pengqi Li$^{1,3}$, Tianhao Wang$^{2,3}$, Lantian Li$^{2*}$, Askar Hamdulla$^{1*}$, Dong Wang$^{3*}$
\thanks{This work was supported by the National Natural Science Foundation of China (NSFC) under Grants No.62171250/6230107/62341607. 
Corresponding author: wangdong99@mails.tsinghua.edu.cn.}
}
\address{$^1$School of Information Science and Engineering, Xinjiang University, China \\
  $^2$School of Artificial Intelligence, Beijing University of Posts and Telecommunications, China \\
  $^3$Center for Speech and Language Technologies, Tsinghua University, China}
\begin{document}
%
\maketitle
\begin{abstract}

Which phonemes convey more speaker traits is a long-standing question, and various perception experiments were conducted with human subjects. 
For speaker recognition, studies were conducted with the conventional statistical models and the drawn conclusions are more or less consistent with the perception results. 
However, which phonemes are more important with modern deep neural models is still unexplored, due to the opaqueness of the decision process. 
This paper conducts a novel study for the attribution of phonemes with two types of deep speaker models that are based on TDNN and CNN respectively, from the perspective of model explanation. 
Specifically, we conducted the study by two post-explanation methods: LayerCAM and Time Align Occlusion (TAO). 
Experimental results showed that: (1) At the population level, vowels are more important than consonants, confirming the human perception studies. 
However, fricatives are among the most unimportant phonemes, which contrasts with previous studies. 
(2) At the speaker level, a large between-speaker variation is observed regarding phoneme importance, indicating that whether a phoneme is important or not is largely speaker-dependent. 

\end{abstract}
\begin{keywords}
\noindent Phonemes, Explanation, Deep speaker model, Speaker recognition
\end{keywords}

\section{Introduction}
\label{sec:intro}

Although deep learning in speaker recognition has achieved great success~\cite{ehsan14,li2017deep,snyder2018x,bai2021speaker}, the contribution of pronunciation units, 
such as phonemes, on speaker recognition performance remains unclear. 
This knowledge lack prevents us from designing more effective architectures and verification schemes. 
More seriously, it prevents a deep understanding of the decision-making mechanism of the model, and thus difficult to tune and control its behavior.

How much speaker-related information is conveyed by each phoneme or phoneme class when humans identify speakers 
has been studied with various perception experiments~\cite{amino2006idiosyncrasy,amino2009speaker}. 
A consistent conclusion is that vowels, nasals, and fricatives are more important than other classes. 
For automatic speaker recognition, a multitude of studies were conducted~\cite{auckenthaler1999improving,kajarekar2001speaker}, 
by constructing individual models for phonemes or phone classes and ranking their performance. 
Nearly all these studies were based on statistical models such as Hidden Markov Model (HMM) or Gaussian Mixture Model (GMM). 
A general conclusion was that vowels and nasals lead to higher performance than other phoneme classes, 
though the contribution of fricatives is not fully agreed~\cite{parris1994discriminative,moez2016phonetic}.




Although the above research is inspiring, the conclusions obtained from perception experiments and with statistical models are not necessarily true for speaker models based on deep neural networks, 
especially with the deep embedding architecture such as the x-vector model and its variants~\cite{snyder2018x}. 
This is because deep speaker models use entire utterances to make decisions, which involves rich context aggregation and complex interaction and competition among phonemes. 
In contrast, both human perception tests and statistical models are generally based on the performance of isolated phonemes.

Very recently, Rafi et al.~\cite{rafi2023relative} investigated the relative contribution of phonemes for an x-vector model using frame-level attention weights, 
and drew conclusions consistent with the perception experiments. 
This study, however, is not fully convincing. 
This is because the attention weights were derived from the last feature layer, making them more `feature importance' rather than `phoneme importance'. 
Moreover, generalizing this approach to other models without the attention layer is not straightforward. 

In this study, we propose a new approach to analyze the contribution of phonemes in deep speaker models. 
This approach employs the recently emerged model-explanation methods, i.e., methods showing how each frame in the input utterance impacts the decision made by the model. 
Specifically, two explanation methods, LayerCAM and Time Align Occlusion (TAO) are employed to analyze each test utterance and generate saliency maps for all the test utterances, 
and the importance of phonemes is derived by aggregating the saliency values of frames belonging to each phoneme. 
This approach can be applied to analyze the behavior of any model, as far as the explanation methods are reliable.


\section{Related work}

The present study is related to model explanation, in particular, model visualization that aims to identify the salient elements of the input data 
that the decision owes to~\cite{samek2017explainable,linardatos2020explainable}. 
For speaker recognition, some studies have used various visualization tools to explain the behavior of deep speaker models~\cite{himawan2019voice,zhou2021resnext,zhang23x_interspeech}. 
A major concern towards this end is whether the visualization tool used is reliable. 
This concern motivated recent research focusing on the reliability of visualization tools~\cite{li2022reliable,li23p_interspeech}. 
A surprising discovery is that blindly using visualization tools can lead to misleading explanations for deep speaker models, and only LayerCAM among the CAM family could provide reliable explanations. 
This work takes advantage of the demonstrated reliability of LayerCAM and employs it to provide utterance-level saliency maps, from which phoneme importances can be derived.

To double confirm the reliability of the results, another visualization tool based on occlusion~\cite{zintgraf2016visualizing,petsiuk2018rise,fong2017interpretable} is also used. 
Considering our goal is to obtain saliency values for frames rather TF (Time-Frequency) bins, a Time Align Occlusion (TAO) is designed, as detailed in the next section. 
We use LayerCAM and TAO to verify each other.

\section{Method}

We use LayerCAM and TAO to analyze two popular deep speaker models, one based on TDNN~\cite{waibel2013phoneme} and the other based on CNN, and the goal is to identify the contribution of different phonemes with these two types of models. 
Firstly the two explanation methods are employed to extract the saliency maps for all the test utterances, and then the importance of each phoneme is obtained by aggregating frame-level saliency, 
referring to phoneme boundaries produced by MFA~\cite{mcauliffe17_interspeech}, a popular forced alignment tool. 
LayerCAM and TAO are briefed below, including some details designed to meet the request of the research purpose.

\subsection{LayerCAM}

LayerCAM~\cite{jiang2021layercam} is a vital tool for visualizing CNN models. 
It constructs a \emph{saliency map} of the same size as the original input, i.e., the Mel spectrum in our case. 
This saliency map shows the important TF regions when a CNN model tries to identify a particular class.

Let $f$ denote the speaker classifier instantiated by a 2D-CNN, and $\theta$ represents its parameters. 
For a given input $x$ from class $c$, the prediction score (posterior probability) for the target class can be computed by a forward pass:
\begin{equation}
\label{eq:pred}
y^c = f_c(x; \theta).
\end{equation}
Secondly, choose the \emph{last} CNN layer that involves a set of activation maps $\{A^k\}$. 
The weight for $k$-th activation map $A^k$ for class $c$ at the location ($i,j$) is defined as the gradient at that location:
\begin{equation}
\label{eq:layer-w}
w_{ij}^{kc} = \text{ReLU } (\frac{\partial y^{c}}{\partial A_{i j}^{k}}).
\end{equation}
Finally, the saliency map for class $c$ is produced as follows:
\begin{equation}
  \label{eq:layer}
  S^c_{ij} =  \text{ReLU } ( \sum_k w_{ij}^{kc} \cdot {A_{i j}^{k}} ).
\end{equation}

\noindent We normalize $S^c_{ij}$ to the range [0,1], following the procedure recommended in~\cite{li2022reliable}.

To obtain a saliency value for each frame, we resize the 2D saliency map to the shape of the input Mel spectrum and 
aggregate the saliency values at all the frequency bins to produce a saliency vector $\xi^c$. 
The value of the $t$-th frame $\xi^c_t$ is:

\begin{equation}
\label{eq:layer}
   \xi^c_t=\sum_{f}(\text{Upsampling}(S^c_{ij}))_{tf}.
\end{equation}

For TDNN, the convolution is one-dimensional over the time axis and there is no downsampling operation. 
The resultant saliency map reduces to a one-dimensional saliency vector $\xi^c$ where each element corresponds to a frame:

\begin{equation}
  \label{eq:layer-2}
  \xi^c_{t} =  \text{ReLU } ( \sum_k w_{t}^{kc} \cdot {A_{t}^{k}} ),
\end{equation}
where $w_t^{kc}$ is computed as follows:

\begin{equation}
\label{eq:layer-w-2}
w_{t}^{kc} = \text{ReLU } (\frac{\partial y^{c}}{\partial A_{t}^{k}}).
\end{equation}

\subsection{Time align occlusion (TAO)}
TAO draws inspiration from the occlusion method described in~\cite{zintgraf2016visualizing,petsiuk2018rise,fong2017interpretable}. 
Firstly it systematically occludes different portions of the input with a perturbation and monitors the output of the classifier.
More specifically, the occlusion is performed by sequentially perturbing a window of input features (e.g., Mel spectrum), with the most common Gaussian blur as the perturbation method.
Each occlusion window covers 7 consecutive frames with a stride of 1 frame. 

By calculating the change in the logit of the target speaker, we can determine the importance of the occluded window by $S_{y}(x) - S_{y}(x_{[x_{i}=blur(x_{i})]})$ 
where $[x_{i}=blur(x_{i})]$ indicates a sample $x$ whose $i$-th component is replaced with the perturbation through Gaussian blur.

We choose the 7-frame occlusion window and the 1-frame stride to match the configuration of the TDNN model, where the receptive field of the neurons in the final layer is 7.
This final leads to a saliency vector $\xi$ where each element represents the saliency value of the corresponding frame. 
Note that TAO is model-agnostic and applies to both TDNN and CNN models without any difference.

\section{Experiment}
\label{sec:exp}

\subsection{DataSet}
\label{sec:dataset}

The Audio-MNIST~\cite{becker2018interpreting} dataset was used in our experiments. 
It comprises 30k recordings of English digits (0-9), and each digit was recorded 50 times by each of the 60 speakers, totaling approximately 9.5 hours. 
This dataset was originally designed to investigate the interpretability of deep neural models on tasks of digit and gender classification. 

Our task here is speaker identification, i.e., identifying the target speaker from a closed set of candidates. 
To support the task, the dataset was equally split into a training set and a test set, each containing all the 60 speakers, and each speaker contributes 25 recordings per digit. 
We concatenate the ten digits (0-9) sequentially to form long utterances. This results in 25 test utterances per speaker, and the total number of utterances is 1,500. 
The special design for the test utterances guarantees that each phoneme shows its contribution in long-span contexts and with full competition with other phonemes. 


\subsection{Deep speaker model}
\label{sec:model}

As mentioned already, our investigation is based on two popular architectures: CNN and TDNN.
For each architecture, we train two models with different complexities.
Details of the four models are shown in Table~\ref{tab:model}.
The notation `$3\times3@32$' means the kernel is (3$\times$3) and the number of channels is 32. 
In all the models, the stride of both the TDNN and CNN layers is set to 1. 
For the CNN models, the max pooling layer performs downsampling, with a window size of $2\times2$ and a stride of 2. 
The models were trained using the Sunine Toolkit\footnote{https://gitlab.com/csltstu/sunine} and the Top-1 accuracy of the identification task is pretty good with any of the four models. 

\begin{table}[!htpb]
  \begin{center}
  \caption{Different models and their Top-1 accuracy.}
  \label{tab:model}
  \scalebox{0.95}{
  \begin{tabular}{lcccc}
    \toprule
    Layer                      & \textbf{TDNN-1} & \textbf{TDNN-2} & \textbf{CNN-3} & \textbf{CNN-4}   \\ 
    \midrule
    Layer-1                    & 5@512           & 5@512           & $3\times3$@32 & $3\times3$@64     \\
                               & ReLU            & ReLU            & ReLU          & ReLU              \\
                               &                 &                 & MaxPool       & MaxPool           \\
   \midrule
   Layer-2                     & 3@512           & 3@1024          & $3\times3$@64 & $3\times3$@64     \\ 
                               & ReLU            & ReLU            & ReLU          & ReLU              \\
                               &                 &                 & MaxPool       & MaxPool           \\
   \midrule                    
   Pooling                     & \multicolumn{4}{c}{Temporal Statistics Pooling (TSP)}                 \\ 
   \midrule
   Dense                       & \multicolumn{4}{c}{128}                                               \\
   Dense                       & \multicolumn{4}{c}{N (60)}                                             \\
   \midrule
   Top-1                       & 100.0\%         & 100.0\%         & 100.0\%        & 100.0\%           \\ 
   \bottomrule
  \end{tabular}}
  \end{center}
\end{table}

\subsection{Phoneme importance distribution (PID)}
\label{sec:exp:pid}

In our test, the phoneme inventory and how the phonemes form the ten digits appearing in Audio-MNIST are shown in Table~\ref{tab:dictionary}.
Note that we labeled the same phoneme in different digits as different variants (e.g., f, f\_2), to account for the context variation. 
Following this scheme, there are 31 context-dependent phonemes in total.

\begin{table}[htpb]
 \begin{center}
   \caption{Phoneme inventory}
   \label{tab:dictionary}
   \scalebox{1.0}{
   \begin{tabular}{cc|cc}
    \toprule
    \textbf{Digit} & \textbf{Phoneme} & \textbf{Digit} & \textbf{Phoneme} \\ 
    \cmidrule(lr){1-1} \cmidrule(lr){2-2} \cmidrule(lr){3-3} \cmidrule(lr){4-4}
    zero             & \textipa{z} \textipa{I} \textipa{\*r}  \textipa{ow}                    & five             & \textipa{f\_}2 \textipa{aj} \textipa{v}                \\
    one              & \textipa{w} \textipa{5} \textipa{n}                   & six              & \textipa{s} \textipa{I} \textipa{k} \textipa{s\_}2            \\
    two              & \textipa{t}$^h$ \textipa{0}              & seven            & \textipa{s\_}3 \textipa{E} \textipa{v\_}2 \textipa{\s{n}}                \\
    three            & \textipa{T} \textipa{\*r\_}2 \textipa{i:}                   & eight            & \textipa{ej} \textipa{P}                 \\
    four             & \textipa{f} \textipa{6} \textipa{\*r\_}3                   & nine             & \textipa{n\_}2 \textipa{aj\_}2 \textipa{n\_}3                 \\
    \bottomrule
  \end{tabular}}
 \end{center}
\end{table}

We define Phoneme Importance Distribution (PID) as a vector that represents the contribution of every phoneme, and it can be computed per utterance or for the whole test set. 
At the utterance level, PID is computed as follows: 
Firstly the MFA tool~\cite{mcauliffe17_interspeech} is used to align the speech frames and the phone sequence to determine the phoneme boundaries (The agreement among four examiners on MFA results is higher than 99\%).
Secondly, compute the importance of each phoneme as follows:
\begin{equation}
\pi_q = \frac{1}{N_q} \sum_{t \in \phi_q} \xi_t,
\end{equation}
where $\xi_t$ is the saliency value of the $t$-th frame computed by either LayerCAM or TAO; 
$q$ is the phoneme index; 
$\phi_q$ is the set of frames belonging to phoneme $q$ (Note that the receptive field covered by these frames contains ONLY the phoneme $q$, without overlapping with other phonemes.);
$N_q$ is the number of frames in $\phi_q$.
Note that each test utterance include all the digits from 0 to 9, so the \textbf{utterance-level PID} vector $\pi$ involves the importance of all the phonemes.

The utterance-level PIDs can be accumulated to produce a \textbf{global PID} by averaging the PIDs of all the test utterances, reflecting the relative importance of different phonemes:
\begin{equation}
\pi^g_q = \frac{1}{N} \sum_n \pi_{nq},
\end{equation}
where $N$ is the number of utterances in the test set, and $\pi_{nq}$ is the utterance-level PID for the $n$-th utterance.

\subsection{Consistency between explanation methods}
\label{sec:cam_TAO}

We first verify if the two explanation methods (LayerCAM and TAO) hold the same opinion regarding phoneme importance. 
Three quantities are computed for verification:

\begin{itemize}
\item Mean correlation on utterance-level saliency vectors:
\[
r1 = \frac{1}{N}\sum_n Corr(\xi^{CAM}_n, \xi^{TAO}_n)
\]
where $\xi^{CAM}_n$ and $\xi^{TAO}_n$ represent the saliency vector of the $n$-th utterance computed by LayerCAM and TAO. $Corr$ denotes the Spearman correlation. 

\item Mean correlation on utterance-level PIDs: 
\[
r2 = \frac{1}{N}\sum_n Corr(\pi^{CAM}_n, \pi^{TAO}_n)
\]
where $\pi^{CAM}_n$ and $\pi^{TAO}_n$ represent PIDs of the $n$-th utterance computed by LayerCAM and TAO. 

\item Correlation on global PIDs:
\[
r3= Corr(\pi^{CAM}_g, \pi^{TAO}_g ).
\]
\end{itemize}

The results are shown in Table~\ref{tab:con}. 
It can be seen that for the TDNN models, LayerCAM and TAO are highly correlated, cross-validating the reliability of the two explanation methods. 
However, for the CNN models, the two methods present very different results. 
We attribute this to the failure of LayerCAM in explaining the CNN models. 
This failure is probably caused by the upsampling and summation operations shown in Eq.\ref{eq:layer}, 
in particular the summation operation that assumes the frame saliency is a simple addition of the saliency of all the frequency bins. 
Accordingly, we will not use LayerCAM to analyze CNN models. 

\begin{table}[!htpb]
  \begin{center}
   \caption{Consistency test between LayerCAM and TAO.}
   \label{tab:con}
   \scalebox{1.0}{
   \begin{tabular}{lcccc}
    \toprule
    \textbf{}     & \textbf{TDNN-1}   & \textbf{TDNN-2}   & \textbf{CNN-3}   & \textbf{CNN-4}    \\
                  \cmidrule(lr){2-2}  \cmidrule(lr){3-3}  \cmidrule(lr){4-4} \cmidrule(lr){5-5}
    $r1$          & 0.864             & 0.869             & 0.141            & 0.323             \\ 
    $r2$          & 0.880             & 0.885             & 0.240            & 0.331            \\ 
    $r3$          & 0.952             & 0.908             & 0.190            & 0.445            \\ 
    \bottomrule
  \end{tabular}}
 \end{center}
\end{table}

\subsection{Consistency between deep speaker models}
\label{sec:model}

The global PIDs can be regarded as an indicator of phone importance assigned by a model.
Therefore, we can examine if different models focus on similar phonemes to make decisions, by computing the correlation among global PIDs produced with different models. 
Table~\ref{tab:con3} presents the results. 
It can be seen that different models view the importance of phonemes similarly, no matter which explanation methods are used. 
This suggests that phoneme importance should be regarded as an intrinsic and model-agnostic property.

\begin{table}[!htpb]
  \centering
  \caption{Consistency test between models. L: LayerCAM, T: TAO; 1: TDNN-1, 2: TDNN-2, 3: CNN-3, 4: CNN-4.}
  \label{tab:con3}
  \scalebox{1.0}{
  \begin{tabular}{c|ccccc}
    \toprule
    \textbf{}       & 2 (T)             & 1 (L)             & 2 (L)             & 3 (T)             & 4 (T)    \\  
    \midrule
    1 (T)               & 0.988      & 0.952      & 0.901      & 0.943      & 0.948       \\
    \midrule
    2 (T)             &            & 0.944      & 0.908      & 0.923      & 0.927        \\ 
    \midrule
    1 (L)            &            &            & 0.961      & 0.931      & 0.950       \\ 
    \midrule
    2 (L)            &            &            &            & 0.879      & 0.923         \\
    \midrule
    3(T)            &            &            &            &            & 0.981         \\ 
    \bottomrule
  \end{tabular}}
\end{table}


\subsection{Phoneme importance}
\label{sec:phoneme}

With the established cross-method and cross-model consistency,
we can estimate phoneme importance using the global PIDs obtained from the four models through the two explanation methods (note that LayerCAM only applies to TDNNs).
The union of the top ten phonemes in the six global PIDs is considered the most important (\textipa{aj}, \textipa{0}, \textipa{aj\_}2, \textipa{5}, \textipa{ej}, \textipa{I}, \textipa{6}, \textipa{E},  \textipa{I\_}2), while the union of the bottom ten phonemes in the six global PIDs is considered the least important (\textipa{k}, \textipa{f}, \textipa{f\_}2, \textipa{s\_}3, \textipa{v}, \textipa{T}, \textipa{P}).
It is clear that the most important phonemes are vowels and the least important phonemes are consonants. 
This is intuitively reasonable and is largely in accordance with the previous findings in the literature. 
However, some observations are unexpected. 
In particular, fricatives such as \textipa{f}, \textipa{s} are among the unimportant phonemes, whereas some previous studies have shown they are speaker-discriminant~\cite{amino2009speaker,parris1994discriminative}. 
These new findings suggest that deep learning models may recognize speakers by focusing on cues different from humans and statistical models. 
We hypothesize that this significant difference is attributed to the competition among phonemes when deep embedding models form utterance-level representations.

\subsection{Speaker variation}
\label{sec:exp:pid1}

Finally, we examined the phoneme importance from the perspective of individual speakers. 
Two quantities are computed with the utterance-level PIDs: Within-speaker correlation $r_w$ and Between-speaker correlation $r_b$, formulated by:
\[
r_w= \frac{1}{S}\sum_s \frac{1}{N_s(N_s-1)}\sum_{i,j, i \ne j} Corr(\pi_{s,i},\pi_{s,j})
\]
\[
r_b= \frac{1}{N}\sum_{S(i) \neq S(j) } Corr(\pi_i,\pi_j)
\]
where $S$, $N_{s}$ denotes the number of speakers and utterances of speaker $s$, respectively, $S(\cdot)$ denotes the speaker label, $\pi_{s,i}$ the PID of the $i$-th utterance of speaker $s$, and $N$ denotes the number of cross-speaker pairs.

Table~\ref{tab:pid-c} shows the results. 
It can be seen that the within-speaker correlation $r_w$ is high, indicating that for a particular speaker, the speaker model always uses the same patterns to distinguish him/her from others. 
In contrast, the between-speaker correlation $r_b$ is much lower, suggesting that different speakers are distinguished by different patterns. 
This is particularly the case for CNN models, where $r_b$ is smaller. 
This might be attributed to the flexibility of 2D CNN in extracting complex TF patterns, thus allowing special and subtle cues to be utilized for each speaker. 
The low $r_b$ means that whether a phoneme is important is largely speaker-dependent, 
and the important phonemes derived in the previous experiment are meaningful just in a statistical sense. 


\begin{table}[!htpb]
  \begin{center}
   \caption{Utterance-level PID consistency}
   \label{tab:pid-c}
   \scalebox{1.0}{
    \begin{tabular}{ccccc}
    \toprule
                &    \multicolumn{4}{c}{Within-Speaker Correlation ($r_w$)}                  \\
    \midrule
    \textbf{}   & \textbf{TDNN-1}  & \textbf{TDNN-2}  & \textbf{CNN-3}   & \textbf{CNN-4}     \\  
                \cmidrule(lr){2-2} \cmidrule(lr){3-3} \cmidrule(lr){4-4} \cmidrule(lr){5-5}
    LayerCAM    & 0.671            & 0.689            & -                & -                  \\ 
    TAO         & 0.696            & 0.701            & 0.553            & 0.548              \\ 
    \midrule
    \midrule
                &    \multicolumn{4}{c}{Between-Speaker Correlation ($r_b$)}                   \\
    \midrule
    \textbf{}   & \textbf{TDNN-1}  & \textbf{TDNN-2}  & \textbf{CNN-3}   & \textbf{CNN-4}      \\  
                \cmidrule(lr){2-2} \cmidrule(lr){3-3} \cmidrule(lr){4-4} \cmidrule(lr){5-5}
    LayerCAM    & 0.360            & 0.353            & -                & -                   \\ 
    TAO         & 0.433            & 0.409            & 0.169            & 0.223               \\ 
    \bottomrule
  \end{tabular}}
 \end{center}
\end{table}

\section{Conclusion}
\label{sec:Conclusion}

Two model explanation methods, LayerCAM and Time Align Occlusion (TAO), were used to analyze the contribution of individual phonemes on two types of deep speaker models, using the Audio-MNIST dataset.
By extensively using the correlation analysis, we first demonstrated that LayerCAM and TAO produce highly consistent results, 
and the examined two deep models, based on TDNN and CNN respectively, show highly consistent behavior. 
The verified consistency among explanation methods and models allows us to compute phoneme importance at the population level and speaker level. 
At the population level, we found that the most important phonemes are vowels and the most unimportant phonemes are consonants. 
The surprising observation is that fricatives are among the unimportant phonemes, in contrast to the results of most previous studies. 
At the speaker level, we found that there is a large speaker variation regarding phoneme importance, indicating that whether a phoneme is important or not is largely speaker-dependent.
Future work involves extending these explanation methods proposed in this paper to additional datasets and languages, to analyze whether the phone importance distributions for speaker recognition possess generalizability. 
Besides, we also contemplate how to utilize these findings as priors in the design of deep speaker models.



\bibliographystyle{IEEEbib}
\bibliography{strings,refs}

\end{document}